**Testing Distinguishability and Determinism in Atoms**

Mark G. Raizen
Dept. of Physics, University of Texas at Austin, Austin, TX 78712

## Abstract

We propose and analyze experiments to test distinguishability and determinism in atoms. As a first step, we consider experiments to test if atoms are distinguishable from each other due to small differences in the nuclear magnetic moment. The idea is to measure the hyperfine splitting for stable isotopes that are confined to ion traps, by accumulating statistics over time with different atoms. The three cases analyzed are strontium-87 with an odd number of neutrons, lutetium-175 with an odd number of protons, and lutetium-176 with an odd number of protons and neutrons. We then consider the possibility of performing the same tests on newly created stable isotopes. We also propose that a time-dependent nuclear magnetic moment may enable deterministic tracking of the life cycle of an isotope, and prediction of age since creation and time until decay of radioisotopes. The case of thulium-170 is considered, it is particularly interesting as a radioisotope with an odd number of protons and neutrons. More generally, we argue that any atomic transition could change in time, either due to the atomic nucleus or to the electron charge distribution. The latter could be tested in a dual atomic clock in a ladder configuration to search for the aging of an electronic transition.

## Introduction

A fundamental assumption of modern physics is that all atoms of a particular isotope and in the same quantum state are identical to each other. A loophole in that argument is the nuclear magnetic moment which is due to the current and spin distributions inside the nucleus [1]. From a theoretical perspective, there is no symmetry or law that requires atoms to have the same nuclear magnetic moment, such as CPT invariance constrained by neutral mesons [2]. In the absence of any symmetry or law, for atoms to be truly indistinguishable would require an unnatural fine-tuning of parameters.

Due to the complexity of non-perturbative quantum chromodynamics, the nuclear magnetic moment and hyperfine splitting in atoms can only be calculated to an accuracy of 0.2% for heavy alkalis [3,4]. In contrast, experimental limits are about eleven orders of magnitude better, so we cannot rely on theory to set any stringent constraints. To test distinguishability of atoms experimentally, one needs an atomic clock to search for possible differences in the nuclear magnetic moment; but in an ensemble, this effect could not be separated from the broadening effects of the measurement. Instead of neutral atoms, single ions must be trapped and their ground state microwave hyperfine splitting (proportional to the nuclear magnetic moment) measured precisely. Due to the very high cost of enriched isotopes, trapped ions are typically stored for long periods, and measurements repeated on the same ion. To test distinguishability of atoms, a different strategy is needed: measure as many ions as possible, at a rate of one per day (in one vacuum chamber). A claim of distinguishability would require measurement statistics at a sufficiently high confidence level. To the best of our knowledge, such an experimental test has not been performed, so the statement that atoms are indistinguishable is just a subjective belief.

In previous papers, we considered the possibility of “atom aging” by measuring a time dependence of a clock transition in radioactive isotopes [5,6]. Atom aging is distinct from distinguishability; one does not imply the other, and the latter could be seen in stable isotopes while the former requires a half-life of a radioisotope. A particular realization of atom aging, not considered before, is that the time variation could deterministically predict the time until decay for an atom.

The question of whether microscopic particles are deterministic was raised in the early days of quantum mechanics [7]. Einstein was convinced that is the case, rejecting a probabilistic version of nature, as expressed in the Copenhagen interpretation of quantum mechanics [8]. Other great scientists have expressed a similar opinion over the years, but the prevailing view today is probabilistic. The issue was considered settled by experiments that observed violation of Bell’s inequalities, which rule out local hidden variables [9]. These experiments relied on spatially separated particles where a quantized value like photon polarization is measured. In cases where there is no spatial separation, the Leggett-Garg inequality was proven for correlations in time [10]. This inequality also required discrete measurables, and violations were observed in several experimental systems such as NMR and Josephson junctions [11].

We consider continuous variables that are produced in composite systems of many interacting particles such as the magnetic moment of the nucleus or the electric quadrupole moment in an atomic transition. These are emergent phenomena, not fundamental or discrete, and depend sensitively on charge or current distributions. To our knowledge, these quantities are not governed by any inequality like Bell or Leggett-Garg, so can serve as local *unhidden* variables. To date, no tests of determinism in many body systems have been considered, and their discovery would require a reformulation of theory. It is not surprising that there are no local hidden variables for fundamental particles such as photons and electrons, but composite systems bring internal complexity not possible for point-like particles. Even a single nucleon, like a free neutron, has three quarks bound together by the strong nuclear force and their charge and current distribution can change in time over its life cycle.

As a first test, we consider possible changes in time in the properties of stable isotopes. Atoms comprising the periodic table were created during nucleosynthesis, the lightest ones during big-bang nucleosynthesis and the heavier ones during stellar and supernova nucleosynthesis [12]. While standard quantum mechanics assumes that stable isotopes are immutable, it is interesting to ask whether any atomic or nuclear property, such as the nuclear magnetic moment, could have changed over billions of years since creation. A nuclear magnetic moment occurs in nuclei with an odd number of neutrons, protons, or both. The exact value can only be calculated to about 10% accuracy, but the atomic hyperfine splitting that is proportional to the nuclear magnetic moment can be measured to an accuracy of one part in $10^{13}$ or better in a single-ion clock with an averaging time of about 1 day [13]. A single-ion optical ion clock in strontium reached an absolute accuracy of $2.3\text{x}10^{-17}$ in recent work [14]. The magnetic moment must change during nucleon rearrangement after formation but the time scale for full equilibration is not known. Today, we can transmute elements by bombarding them with particles such as neutrons in a nuclear reactor or protons in a cyclotron, which are used to produce radioisotopes for basic research and applications. We propose to use nuclear transmutation to produce *stable* isotopes, thereby recreating the initial conditions during nucleosynthesis. The nuclear spin is determined by the number of protons (Z) and neutrons (N) in the nucleus, and there are four possibilities: even-Z & even-N, with zero nuclear spin and magnetic moment; even-Z & odd-N; odd-Z & even-N; odd-Z & odd-N. The last three combinations result in a non-zero nuclear spin and magnetic moment, and the odd-Z & odd-N case is very rare and is associated with weaker nuclear binding forces. We focus on isotopes that can be used for trapped-ion microwave clocks due to their favorable optical transitions that allow laser cooling and state-resolved detection. A radio frequency (RF) linear Paul ion trap could be used to enable measurements of a string of newly created stable isotopes, where each ion could be measured separately in succession [15-16]. This would search for a possible variation of the nuclear magnetic moment between atoms, making them distinguishable.

**Stable Isotopes with Even-Z & Odd-N**

Several even-Z & odd-N stable isotopes can be used as hyperfine ion clocks, including Ca-43, Sr-87, and Yb-171. We focus here on Sr-87 with a nuclear spin of 9/2 and a magnetic moment of -1.093603 $\mu$N (unit of nuclear magneton). A Grotrian diagram of Sr-87$^+$ is shown in Fig. 1. The ground state F=4 to F=5 hyperfine splitting is approximately 5 GHz, which can be measured using microwave-optical double resonance spectroscopy [13]. The ion's state is first prepared by optically pumping to the F=4 state with a laser near 422 nm tuned to the upper $P_{1/2}$ state. The microwave drive is applied as a Ramsey pulse sequence, and the upper F=5 hyperfine state is detected with a laser near 408 nm as a cycling transition with the upper $P_{3/2}$ state. To load a Sr-87 ion in an RF linear Paul trap (or multiple ions in a linear trap), neutral Sr-87 can be vaporized near the trap, and photo-ionized with two lasers near 461 nm and 405 nm that are focused to the center of the trap [17]. To produce new Sr-87, one can enrich Sr-86, a stable isotope with a natural abundance of 9.86%. This can be accomplished by electromagnetic separation [18], atomic vapor laser isotope separation (AVLIS) [19], magnetically activated and guided isotope separation (MAGIS) [20], or laser ionization by Multi-Resonant Laser Isotope Separation (MRLIS), a new efficient method that was recently proposed [21]. In the case of MRLIS, the desired isotope would be excited continuously with one laser near 461 nm and ionized by a second laser near 405 nm inside a large mode-volume resonant cavity. The suppression of unwanted isotopes, especially Sr-87, would be in the range of $10^3$-$10^5$, depending on the method used. This is not sufficient for the current purpose, where a suppression factor of at least $10^8$ is needed to avoid contamination of the new isotopes to be produced. A second pass with a collimated atomic beam using a resonant pushing beam to remove residual Sr-87 atoms from the neutral atom stream could reach the required purity. We have simulated the neutron transmutation of Sr-86 in a 1 MW TRIGA reactor, using a thermal neutron flux of 1.0x $10^{13}$ neutrons/s.cm$^2$ in the Central Thimble (details for this simulation and the others listed below are provided in Supplementary Material). This process would produce a nuclear isomer Sr-87m, which decays with a half-life of 2.8 hours to Sr-87. The production of new Sr-87 atoms is shown in Fig. 2 as a function of irradiation time and displays linear growth for a period of 30 days. A target of 20 g of enriched Sr-86 would produce 0.1 mg of newly formed Sr-87 in five days of irradiation, which could be separated from the target using MRLIS with nearly perfect recovery efficiency.

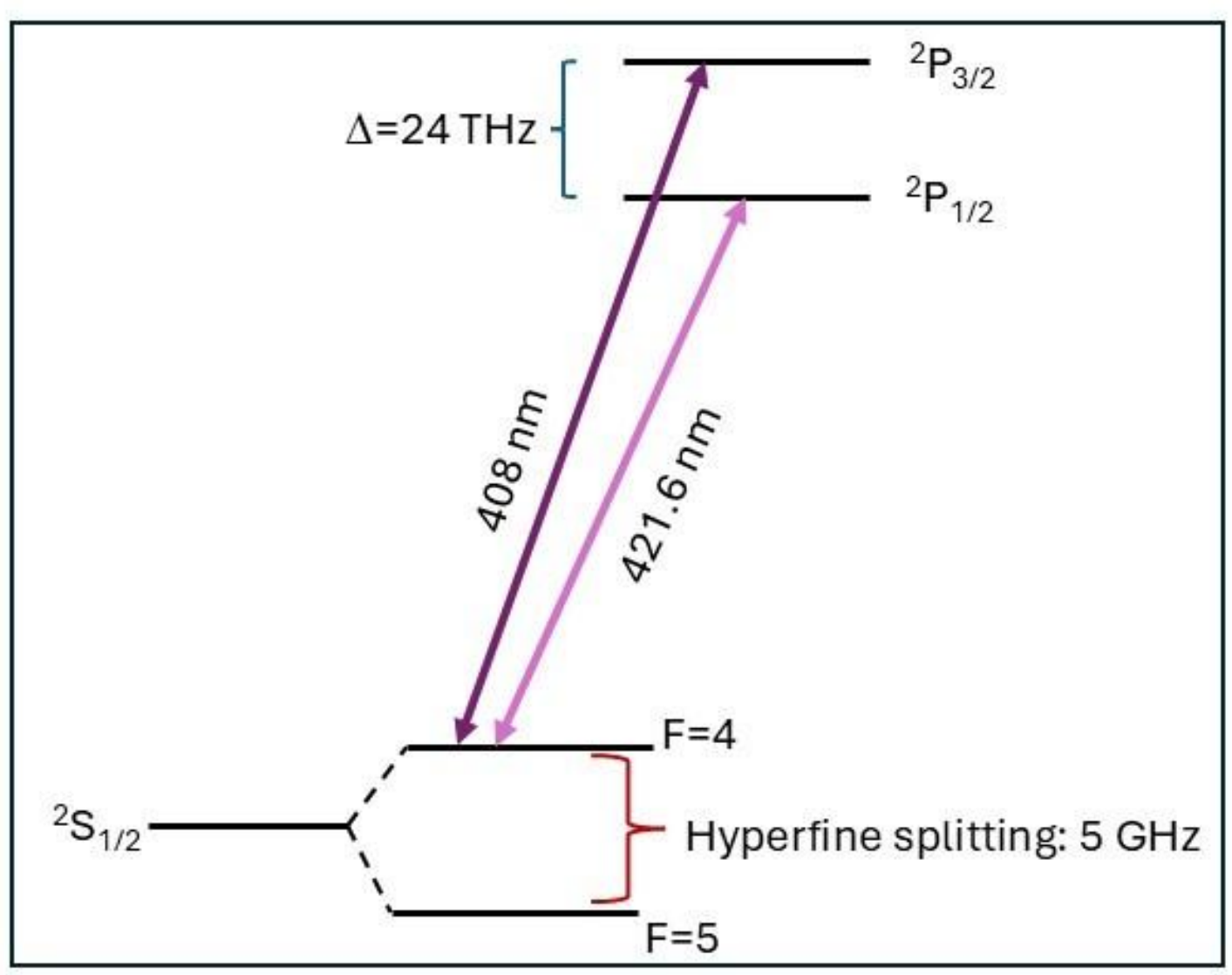

**Fig. 1.** Simplified Grotrian diagram for Sr-87$^+$.

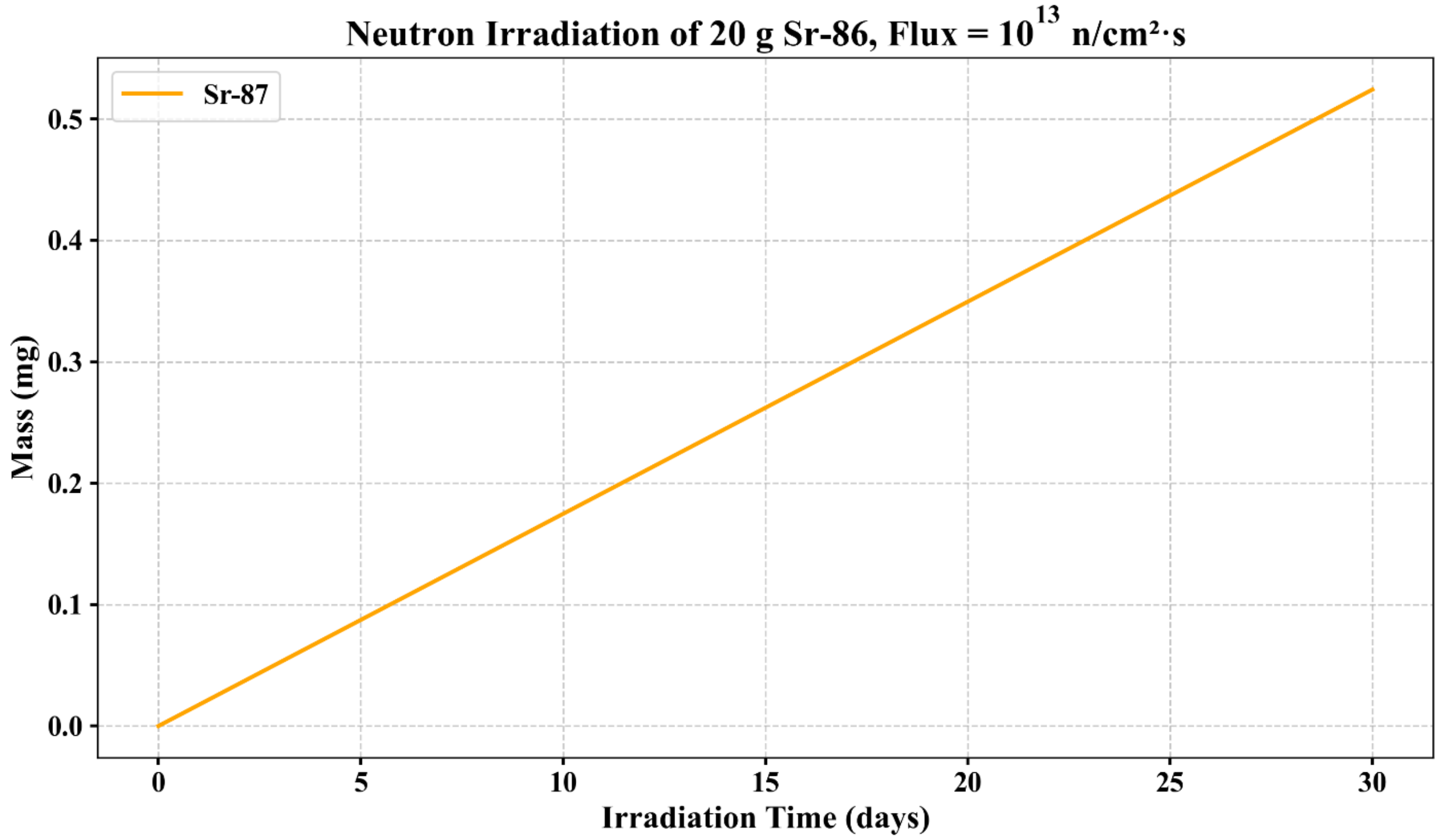

**Fig. 2.** Numerical simulation of production of Sr-87 (orange line) in a nuclear reactor as a function of irradiation time of Sr-86. The amount of Sr-88 produced by double-neutron capture is negligible.

## Stable Isotopes with Odd-Z & Even-N

The only odd-Z & even-N stable isotopes that are suitable as hyperfine ion clocks are Al-27 and Lu-175. We focus here on Lu-175 with nuclear spin of 7/2 and magnetic moment of 2.2327 $\mu$N. To produce new Lu-175, the starting point is enrichment of the stable isotope Yb-174, with a natural abundance of around 32%. This target would be irradiated in a nuclear reactor to produce Yb-175m by a (n,$\gamma$) reaction, decaying in 68 ms to the radioisotope Yb-175 with a half-life of about 4.2 days, decaying into stable Lu-175. Highly enriched Yb-174 is commercially available. A numerical simulation of production of Yb-175 was performed, and the results are shown in Fig. 3. The resulting Lu-175 can be removed from the Yb-174 target by radiochemistry, or by MRLIS. In the latter case, Lu-175 atoms would be excited continuously with one laser near 452 nm and ionized by a second laser near 462 nm inside a large mode-volume resonant cavity. Lutetium ions are already used for ultra-precise atomic clocks [22], and a comparison between new Lu-175 and natural Lu-175 is straightforward.

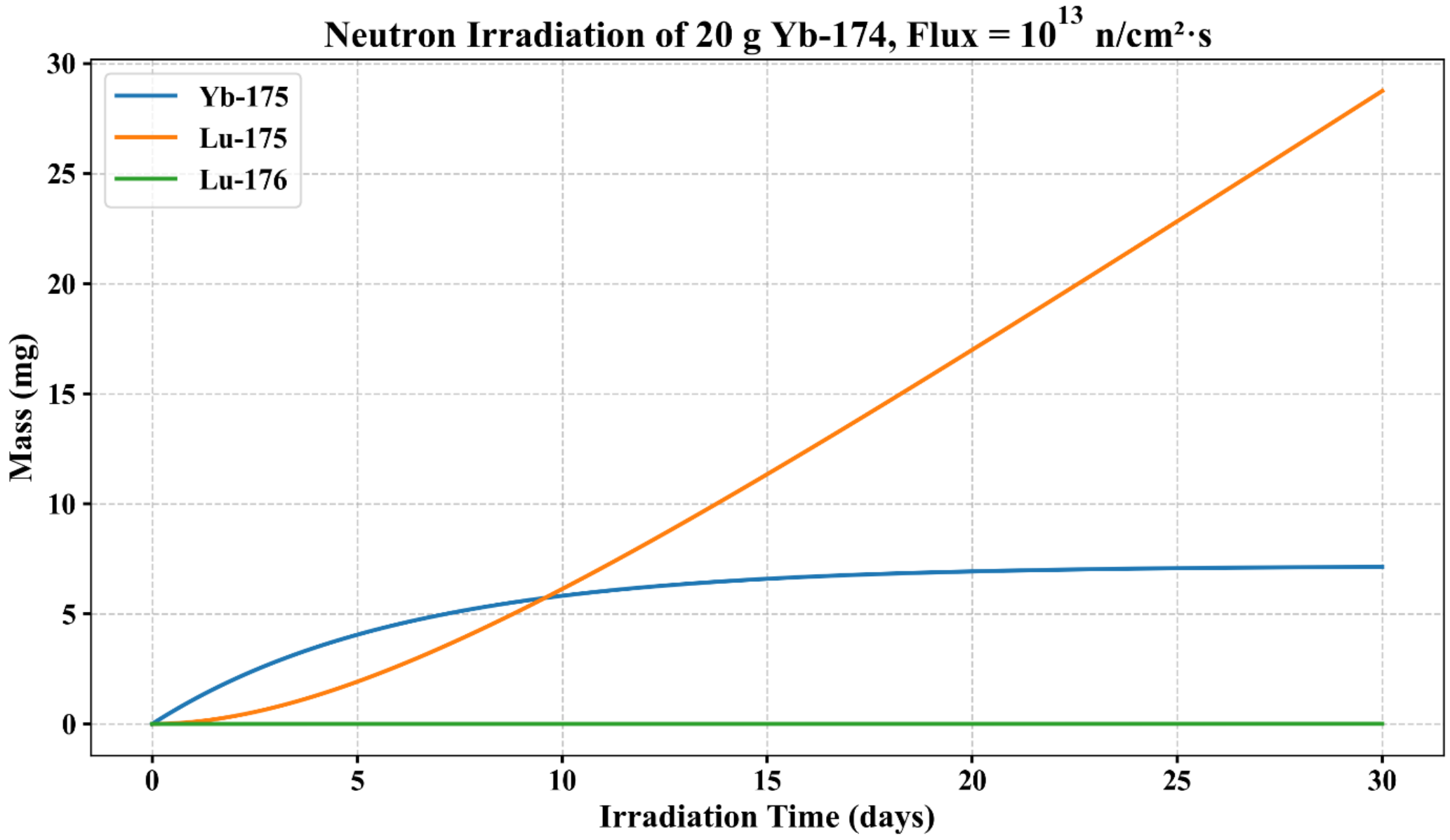

**Fig. 3.** Numerical simulation of production of Lu-175 (orange line) in a nuclear reactor as a function of irradiation time of Yb-174. Co-production of Yb-175 is shown (blue line). The amount of Lu-176 produced by double-neutron capture (green line) is negligible.

## Stable Isotopes with Odd-Z & Odd-N

The only odd-Z & odd-N isotope that is suitable as a hyperfine ion clock is Lu-176, with nuclear spin of 7 and magnetic moment of 3.1692 μN. Although this isotope is not strictly stable, its half-life is 3.76 x $10^{10}$ years, much longer than the lifetime of the universe. To produce new Lu-176, the starting point is enrichment of the stable isotope Lu-175, with a natural abundance of around 97.2%. It can be enriched by MRLIS using the same method described above, and the residual Lu-176 could be pushed away with a resonant laser near 452 nm. A numerical simulation of production of Yb-176 was performed, and the results are shown in Fig. 4. One concern could be the absorption of a second neutron by Lu-176, due to the very large cross section (over 2,000 barns) for this process. We find that this is not a concern, at least for the TRIGA reactor flux and for an irradiation time of 30 days. The resulting Lu-176 could be removed from the Lu-175 by MRLIS and collected for later use.

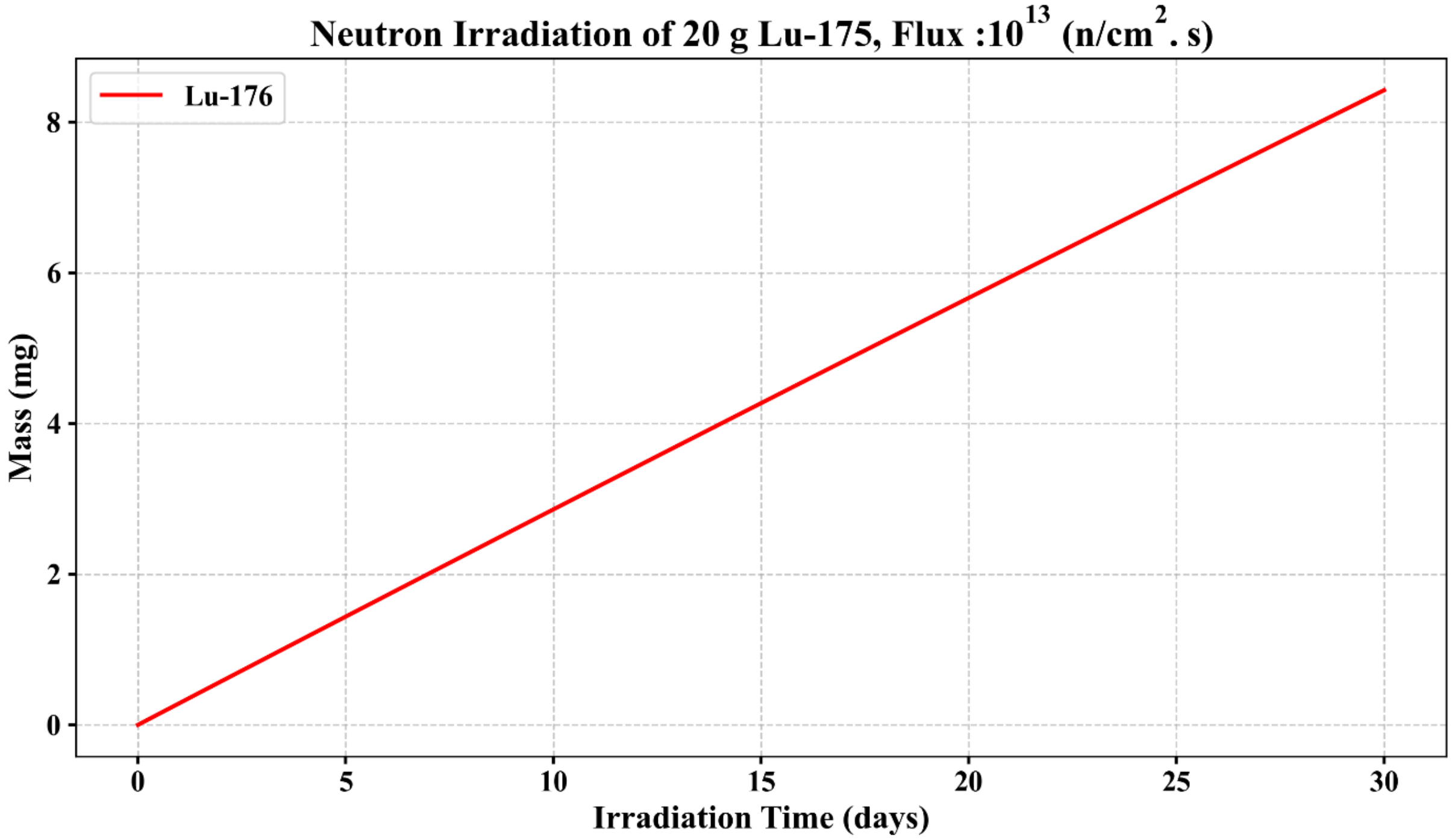

**Fig. 4.** Numerical simulation of production of Lu-176 in a nuclear reactor as a function of irradiation time of Lu-175. The amount of Lu-177 produced by double-neutron capture is negligible.

## Radioisotopes with Odd-Z & Odd-N

There are several odd-Z & odd-N radioisotopes that are suitable as hyperfine ion clocks, we identified thulium-170 (Tm-170) as the most promising case. It has a half-life of about 128.6 days, undergoing beta decay to stable Yb-170. Tm-170 has 69 protons and 101 neutrons, with a magnetic moment of 0.2476 Nm. There is only one stable isotope of thulium, Tm-169, so it can be directly irradiated in a reactor. A numerical simulation of production of Tm-170 was performed, and the results are similar to the case of Lu-175 (Fig. 3).

The atomic structure of Tm-170 is amenable to MRLIS separation from the target Tm-169. One laser near 283.5 nm can excite the atoms from the ground state, with a rate of $1.13 \times 10^6$ $s^{-1}$, and photoionization accomplished with a second laser near 684 nm.

Likewise, the ion Tm-170+ is amenable to laser cooling with a laser near 346 nm, and operation of a microwave hyperfine clock of the ground state with fluorescence detection.

## General Case of Atomic Transitions

The previous discussion was limited to isotopes with a non-zero magnetic moment, which are relatively easy to probe using microwave-optical double resonance spectroscopy of the ground hyperfine state. Regardless of magnetic moment, every atomic transition is affected by the nucleus, known as the isotope shift. It has two origins, a mass shift and a field shift. The latter is due to the nuclear charge distribution. To test for a time variation in the isotope shift with trapped ions, one could use an optical forbidden transition, such as electric quadrupole. This requires an ultra-stable laser, locked to a reference cavity, two ion traps for comparison of isotopes, and an acousto-optic or electro-optic modulator to shift the laser frequency to each ion resonance. Such tests could be done by metrology laboratories using trapped atomic ions.

We next consider the situation where the atomic nucleus is stable, but the atomic electronic state is unstable. Can one predict when an excited atomic state will decay? The theory of quantum jumps predicted a random process [23]. The observation of quantum jumps in single trapped ions confirmed the theory, and the signals were completely random “telegraph signals” [24-25].

To observe a possible aging effect, we propose the following atomic ladder consisting of a ground state g, and two excited states, $e_1$ and $e_2$ as shown in Fig. 5. We assume that the first transition, g to $e_1$ is a long-lived clock transition (Clock 1). The second, $e_1$ to $e_2$ is also a clock transition, but with a much shorter lifetime (Clock 2). State $e_2$ can be detected by an allowed dipole transition to an upper state $e_3$. The experiment would be done on a single trapped ion, and initially it would be excited to state $e_1$ by a microwave source, laser or discharge. One would repeatedly probe the second clock transition until the ion decays from state $e_1$ to g. The measurements would not be too frequent so as not to disturb state $e_1$, as in the Quantum Zeno effect [26-27]. An aging effect could be seen in the frequency of Clock 2 due to a changing electric quadrupole moment.

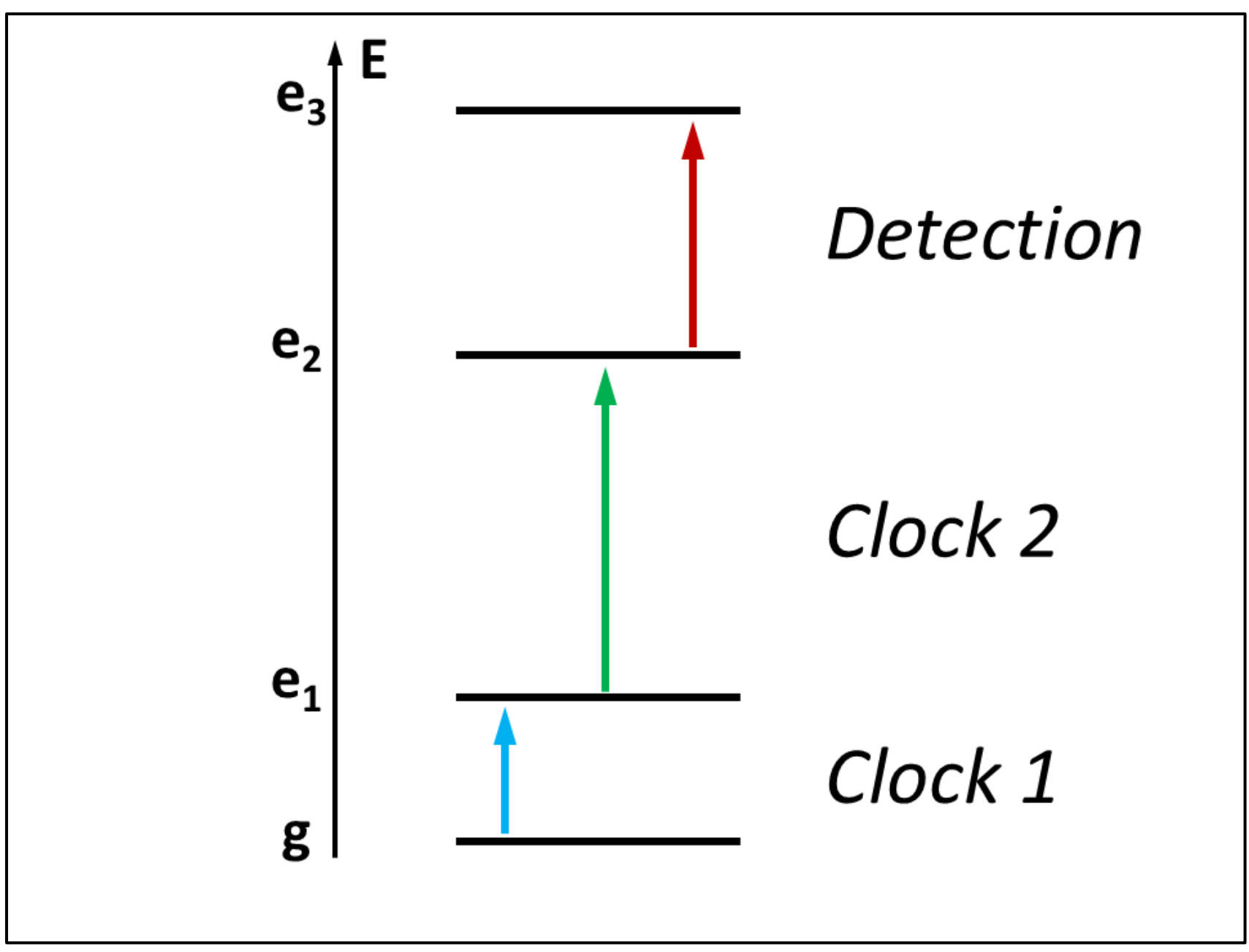

**Fig. 5.** Schematic energy levels and transitions for a model atom that could test atom aging.

As a physical example, this could be realized with a Sr-87 ion clock in a linear RF Paul trap [28]. Clock 1 would be the ground state hyperfine splitting near 5 GHz. Clock 2 would be an electric quadrupole near 674 nm [14], and both have been realized separately.

## Conclusions

In summary, we propose and analyze tests for distinguishability and determinism in atoms where the nuclear magnetic moment can be measured with very high precision. It is a very general picture, applicable to the smallest composite particle like a free neutron, or to larger composite particles like heavy nuclei, atoms, or even molecules. Precision measurements of the type proposed here could potentially answer fundamental questions on the nature of the microscopic world.

## Acknowledgements

I thank Ahmed Helal for his help with the manuscript and the supplementary material. I also thank Aaron Barr, and Henry Chance for their help with the manuscript, and Dmitry Budker for helpful discussions. This work is supported by the Novo Nordisk Foundation and the Sid W. Richardson Foundation.